# Solution of the Dirac equation by separation of variables in spherical coordinates for a class of three-parameter non-central electromagnetic potential

A. D. Alhaidari

*Physics Department, King Fahd University of Petroleum & Minerals, Dhahran 31261, Saudi Arabia*
e-mail: haidari@mailaps.org

We consider the three-dimensional Dirac equation in spherical coordinates with coupling to static electromagnetic potential. The space components of the potential have angular (non-central) dependence such that the Dirac equation is separable in all coordinates. We obtain exact solutions for the case where the potential satisfies the Lorentz gauge fixing condition and its time component is the Coulomb potential. The relativistic energy spectrum and corresponding spinor wavefunctions are obtained. The Aharonov-Bohm and magnetic monopole potentials are included in these solutions.

## I. INTRODUCTION

Dirac equation for spin $\frac{1}{2}$ particle in four dimensional Minkowski space-time is a relativistically covariant first order linear differential equation for a four-component wavefunction ("spinor") $\psi$. For a free structureless particle it reads $\left(i\hbar\gamma^\mu\partial_\mu - \mathcal{M}c\right)\psi = 0$, where $\mathcal{M}$ is the rest mass of the particle and $c$ the speed of light. The summation convention over repeated indices is used. That is, $\gamma^\mu\partial_\mu = \gamma^0\partial_0 + \vec{\gamma}.\vec{\partial} = \gamma^0\frac{\partial}{c\partial t} + \vec{\gamma}.\vec{\nabla}$, where $\{\gamma^\mu\}_{\mu=0}^3$ are four constant square matrices satisfying the anticommutation relation $\{\gamma^\mu,\gamma^\nu\} = \gamma^\mu\gamma^\nu + \gamma^\nu\gamma^\mu = 2\mathcal{G}^{\mu\nu}$, where $\mathcal{G}$ is the metric of Minkowski space-time which is equal to $\mathrm{diag}(+ - - -)$. A four-dimensional matrix representation that satisfies this relation is taken as follows:

$$\gamma^0 = \begin{pmatrix} I & 0 \\ 0 & -I \end{pmatrix}, \quad \vec{\gamma} = \begin{pmatrix} 0 & \vec{\sigma} \\ -\vec{\sigma} & 0 \end{pmatrix}, \tag{1.1}$$

where $I$ is the 2×2 unit matrix and $\vec{\sigma}$ are the three 2×2 hermitian Pauli spin matrices:

$$\sigma_1 = \begin{pmatrix} 0 & 1 \\ 1 & 0 \end{pmatrix}, \quad \sigma_2 = \begin{pmatrix} 0 & -i \\ i & 0 \end{pmatrix}, \quad \sigma_3 = \begin{pmatrix} 1 & 0 \\ 0 & -1 \end{pmatrix}. \tag{1.2}$$

We adopt the atomic units, $\hbar = \mathcal{M} = 1$, in which the relativistic parameter is the Compton wavelength $\lambdabar = \hbar/\mathcal{M}c = 1/c$ and the Dirac equation reads $\left(i\gamma^\mu\partial_\mu - \lambdabar^{-1}\right)\psi = 0$. Our choice of units over the usual relativistic units, where $\hbar = c = 1$, is made to allow us to take the nonrelativistic limit $c \to \infty$ ($\lambdabar \to 0$) which is not possible (for some types of interactions) in the latter units since $c = 1$. Next, we let the Dirac spinor be charged and coupled to the four component electromagnetic potential $A_\mu = (A_0, c\vec{A})$. Gauge invariant coupling, which is accomplished by the "minimal substitution" $\partial_\mu \to \partial_\mu + i\lambdabar A_\mu$, trans-forms the free Dirac equation above to $\left[i\gamma^\mu(\partial_\mu + i\lambda A_\mu) - \lambda^{-1}\right]\psi = 0$. When written in details, it reads as follows

$$i\lambda \frac{\partial}{\partial t}\psi = \left(-i\vec{\alpha}\cdot\vec{\nabla} + \vec{\alpha}\cdot\vec{A} + \lambda A_0 + \lambda^{-1}\beta\right)\psi, \tag{1.3}$$

where $\vec{\alpha}$ and $\beta$ are the hermitian matrices

$$\vec{\alpha} = \gamma^0\vec{\gamma} = \begin{pmatrix} 0 & \vec{\sigma} \\ \vec{\sigma} & 0 \end{pmatrix}, \quad \beta = \gamma^0 = \begin{pmatrix} I & 0 \\ 0 & -I \end{pmatrix}. \tag{1.4}$$

For time independent potentials, equation (1.3) gives the following matrix representation of the Dirac Hamiltonian (in units of $mc^2 = \lambda^{-2}$)

$$\mathcal{H} = \begin{pmatrix} 1 + \lambda^2 A_0 & -i\lambda\vec{\sigma}\cdot\vec{\nabla} + \lambda\vec{\sigma}\cdot\vec{A} \\ -i\lambda\vec{\sigma}\cdot\vec{\nabla} + \lambda\vec{\sigma}\cdot\vec{A} & -1 + \lambda^2 A_0 \end{pmatrix}. \tag{1.5}$$

Thus the energy eigenvalue wave equation reads $(\mathcal{H} - \varepsilon)\psi = 0$, where $\varepsilon$ is the relativistic energy which is real and measured in units of $mc^2$.

Now, the Dirac equation is invariant under the usual abelian electromagnetic gauge transformation $A_\mu \to A_\mu + \partial_\mu \zeta$, $\psi \to e^{-i\zeta}\psi$, where $\zeta$ is a real space-time scalar function. Consequently, the contribution of the off-diagonal term $\lambda\vec{\sigma}\cdot\vec{A}$ could be eliminated ("gauged away"). Therefore, we choose to fix this gauge degree of freedom by requiring that the electromagnetic potential satisfy the relativistically invariant constraint $\partial_\mu A^\mu = 0$, which is referred to as the "Lorentz gauge." For time-independent potentials this constraint reduces to the "radiation gauge," $\vec{\nabla}\cdot\vec{A} = 0$. It should also be noted that the theory still contains residual gauge freedom despite the gauge fixing. However, these gauge fields are free fields and decouple from the physical fields even in the presence of interactions since they satisfy the wave equation $\Box^2 \zeta = 0$, where $\Box = \partial_\mu \partial^\mu$. Fixing the gauge, or in better terms restricting the gauge, is important as could also be seen from the prospective of quantum field theory. In that theory one of the most important objects in scattering calculation, which is performed using the tools of the Feynman diagrams, is the particle propagator (the two-point Green's function). This propagator is the inverse of the wave operator. To define such an inverse it is required that the null space (kernel) of the wave operator be small enough such that it is possible to find boundary conditions that could be used to define a propagator which is nonsingular in a large enough region of configuration (or momentum) space. Unrestricted gauge invariance of the wave operator means that there exists a large space of solutions (the gauge modes) that solves the wave equation trivially. That is, the kernel (null space) of the wave operator is large and one may not be able to define its inverse, the propagator. That's why we fix, or restrict, the gauge degree of freedom.

Solutions of the Dirac equation in one-, two-dimension and in three dimensions with spherical symmetry and for different kinds of configuration of the electromagnetic potential and for different choices of gauge have been reported extensively in the literature. For a review, one may consult, for example, the book by Thaller [1] and references therein. However, little work has been done in three dimensions where the electromagnetic potential has angular variations and the solution requires separation of variables [2]. Moreover, most of the work on separation of variables in the Dirac equation concentrated on separation in curved space-times with gravitational coupling and in the absence of the electromagnetic potential with the exception of very few.

Prominent and sustained contributions to the field are made by Villalba, Shishkin, Bagrov, Cabos, Obukhov, del Castillo and others. For the latest publications with citation to earlier original work, one may consult the papers listed in [3]. The relativistic Aharonov-Bohm effect and the relativistic extension of the magnetic monopole potential are examples of such applications.

In this article we set out to present a systematic and intuitive approach for the separation of variables in spherical coordinates for the Dirac equation with coupling to static non-central electromagnetic potential. In the following section, we construct the solution space for the angular component of the Dirac equation. In Sec. III, the Lorentz gauge fixing condition is imposed and we obtain the solution of the radial equation for the case where the time component of the electromagnetic potential is the Coulomb. The relativistic energy spectrum associated with this non-central electromagnetic potential is obtained. In Sec. IV, we show that the solution of the spherically symmetric problem is a special case of our findings. Moreover, we obtain the nonrelativistic limit and verify that it reproduces results already found in the literature.

## II. SEPARATION OF VARIABLES IN SPHERICAL COORDINATES AND SOLUTION OF THE ANGULAR DIRAC EQUATION

We let the Dirac spinor be charged and coupled to the time-independent electromagnetic potential with the following components in spherical coordinates

$$A_0(\vec{r}) = V(r), \quad A_r(\vec{r}) = W_r(r), \quad A_\theta(\vec{r}) = W_\phi(\phi)/r\sin\theta, \quad A_\phi(\vec{r}) = W_\theta(\theta)/r, \quad (2.1)$$

where $V$, $W_r$, $W_\theta$, and $W_\phi$ are real potential functions. If $W_r \sim r^{-2}$ or $W_r = 0$, then this four-potential satisfies the Lorentz gauge fixing condition, $\partial_\mu A^\mu = 0$. Writing the spinor wavefunction as $\psi(\vec{r}) = \begin{pmatrix} if_+(\vec{r}) \\ f_-(\vec{r}) \end{pmatrix}$, then the action of the Dirac Hamiltonian (1.5) on the four-component spinor $\begin{pmatrix} f_+ \\ f_- \end{pmatrix}$ is represented by

$$\mathcal{H} = \mathcal{H}_0 + \lambda \vec{\sigma} \cdot \hat{r} \mathcal{H}_r + \lambda \frac{\vec{\sigma} \cdot \hat{\theta}}{r} \mathcal{H}_\theta + \lambda \frac{\vec{\sigma} \cdot \hat{\phi}}{r\sin\theta} \mathcal{H}_\phi, \quad (2.2)$$

where $(\hat{r}, \hat{\theta}, \hat{\phi})$ are the unit vectors in spherical coordinates and

$$\mathcal{H}_0 = \begin{pmatrix} 1 + \lambda^2 V & 0 \\ 0 & -1 + \lambda^2 V \end{pmatrix}, \quad \mathcal{H}_r = \begin{pmatrix} 0 & -\partial_r - iW_r \\ \partial_r + iW_r & 0 \end{pmatrix}, \quad (2.3a)$$

$$\mathcal{H}_\theta = \begin{pmatrix} 0 & -\partial_\theta - iW_\phi/\sin\theta \\ \partial_\theta + iW_\phi/\sin\theta & 0 \end{pmatrix}, \quad (2.3b)$$

$$\mathcal{H}_\phi = \begin{pmatrix} 0 & -\partial_\phi - iW_\theta \sin\theta \\ \partial_\phi + iW_\theta \sin\theta & 0 \end{pmatrix}. \quad (2.3c)$$

Square integrability (with respect to the measure $d^3\vec{r} = r^2 dr \sin\theta d\theta d\phi$) and the boundary conditions require that $\psi(\vec{r})$ satisfies: $r\psi(r)\big|_{\substack{r=0 \\ r\to\infty}} = 0$, $\sqrt{\sin\theta}\,\psi(\theta)\big|_{\substack{\theta=0 \\ \theta=\pi}}$ is finite, and $\psi(\phi) = \psi(\phi + 2\pi)$. To simplify the construction of the solution, we look for a local 2×2 similarity transformation $\Lambda(\vec{r})$ that maps the spherical projection of the Pauli matrices ($\vec{\sigma}\cdot\hat{\theta}, \vec{\sigma}\cdot\hat{\phi}, \vec{\sigma}\cdot\hat{r}$) into their canonical Cartesian representation ($\sigma_1, \sigma_2, \sigma_3$), respectively. That is,

$$\Lambda^{-1}\vec{\sigma}\cdot\hat{\theta}\,\Lambda = \sigma_1, \ \Lambda^{-1}\vec{\sigma}\cdot\hat{\phi}\,\Lambda = \sigma_2, \ \Lambda^{-1}\vec{\sigma}\cdot\hat{r}\,\Lambda = \sigma_3. \tag{2.4}$$

Other rearrangement or permutations of the $\sigma_i$'s on the right are equivalent, differing only by a unitary transformation. A 2×2 matrix that satisfies (2.4) is

$$\Lambda(\vec{r}) = \lambda(\vec{r}) e^{-\frac{i}{2}\sigma_3\phi} e^{-\frac{i}{2}\sigma_2\theta}, \tag{2.5}$$

where $\lambda(\vec{r})$ is a 1×1 real function and the exponentials are 2×2 unitary matrices. The transformed wavefunction, which we write as $\chi = \begin{pmatrix} g_+ \\ g_- \end{pmatrix}$, has the two-component spinors $g_\pm = \Lambda^{-1} f_\pm$, whereas the Dirac Hamiltonian (2.2) gets mapped into

$$H = H_0 + \lambda\sigma_3 H_r + \lambda\frac{\sigma_1}{r}H_\theta + \lambda\frac{\sigma_2}{r\sin\theta}H_\phi, \tag{2.6}$$

where

$$H_0 = \mathcal{H}_0, \ H_r = \begin{pmatrix} 0 & -\Lambda^{-1}\partial_r\Lambda - iW_r \\ \Lambda^{-1}\partial_r\Lambda + iW_r & 0 \end{pmatrix}, \tag{2.7a}$$

$$H_\theta = \begin{pmatrix} 0 & -\Lambda^{-1}\partial_\theta\Lambda + \sigma_3 W_\theta \\ \Lambda^{-1}\partial_\theta\Lambda - \sigma_3 W_\theta & 0 \end{pmatrix}, \tag{2.7b}$$

$$H_\phi = \begin{pmatrix} 0 & -\Lambda^{-1}\partial_\phi\Lambda - \sigma_3 W_\phi \\ \Lambda^{-1}\partial_\phi\Lambda + \sigma_3 W_\phi & 0 \end{pmatrix}, \tag{2.7c}$$

and

$$\Lambda^{-1}\partial_r\Lambda = \partial_r + \tfrac{\lambda_r}{\lambda}, \ \Lambda^{-1}\partial_\theta\Lambda = \partial_\theta + \tfrac{\lambda_\theta}{\lambda} - \tfrac{i}{2}\sigma_2, \tag{2.8a}$$

$$\Lambda^{-1}\partial_\phi\Lambda = \partial_\phi + \tfrac{\lambda_\phi}{\lambda} + \tfrac{i}{2}(\sigma_1\sin\theta - \sigma_3\cos\theta), \tag{2.8b}$$

with $\lambda_k = \partial_k\lambda$. This gives the following components of the Dirac Hamiltonian in (2.6)

$$H_r = \begin{pmatrix} 0 & -\partial_r - \tfrac{\lambda_r}{\lambda} - \tfrac{1}{r} - iW_r \\ \partial_r + \tfrac{\lambda_r}{\lambda} + \tfrac{1}{r} + iW_r & 0 \end{pmatrix}, \tag{2.9a}$$

$$H_\theta = \begin{pmatrix} 0 & -\partial_\theta - \tfrac{\lambda_\theta}{\lambda} - \tfrac{\cos\theta}{2\sin\theta} + \sigma_3 W_\theta \\ \partial_\theta + \tfrac{\lambda_\theta}{\lambda} + \tfrac{\cos\theta}{2\sin\theta} - \sigma_3 W_\theta & 0 \end{pmatrix}, \tag{2.9b}$$

$$H_\phi = \begin{pmatrix} 0 & -\partial_\phi - \tfrac{\lambda_\phi}{\lambda} - \sigma_3 W_\phi \\ \partial_\phi + \tfrac{\lambda_\phi}{\lambda} + \sigma_3 W_\phi & 0 \end{pmatrix}. \tag{2.9c}$$

Thus, hermiticity of the Dirac Hamiltonian (2.6) requires that

$$\lambda_\phi = 0, \ \tfrac{\lambda_r}{\lambda} + \tfrac{1}{r} = 0, \ \tfrac{\lambda_\theta}{\lambda} + \tfrac{\cos\theta}{2\sin\theta} = 0, \tag{2.10}$$

giving $\lambda(\vec{r}) = 1/r\sqrt{\sin\theta}$. It is interesting to note that $1/\lambda^2$ turns out to be the integration measure in spherical coordinates. We could have eliminated the $\lambda$ factor in the definition of $\Lambda$ in (2.5) by proposing that the new spinor wavefunction $\chi$ be defined by $\psi(\vec{r}) = \frac{1}{r\sqrt{\sin\theta}}\chi(\vec{r})$. In that case, the transformation matrix $\Lambda$ becomes only $e^{-\frac{i}{2}\sigma_3\phi}e^{-\frac{i}{2}\sigma_2\theta}$, which is unitary. However, making the presentation as above gave us a good chance to show (in a different approach) why is it that one customarily takes the radial component of the wavefunction in spherical coordinates to be proportional to $1/r$ and sometimes the angular component to be proportional to $1/\sqrt{\sin\theta}$. Finally, we obtain the following complete Dirac equation $(H-\varepsilon)\chi = 0$:

$$\left[\begin{pmatrix} 1+\lambda^2 V-\varepsilon & -\lambda\sigma_3(\partial_r+\mathrm{i}W_r) \\ \lambda\sigma_3(\partial_r+\mathrm{i}W_r) & -1+\lambda^2 V-\varepsilon \end{pmatrix} + \frac{\lambda}{r}\begin{pmatrix} 0 & -\sigma_1\partial_\theta-\mathrm{i}\sigma_2 W_\theta \\ \sigma_1\partial_\theta+\mathrm{i}\sigma_2 W_\theta & 0 \end{pmatrix}\right.$$
$$\left. + \frac{\lambda}{r\sin\theta}\begin{pmatrix} 0 & -\sigma_2\partial_\phi-\mathrm{i}\sigma_1 W_\phi \\ \sigma_2\partial_\phi+\mathrm{i}\sigma_1 W_\phi & 0 \end{pmatrix}\right]\begin{pmatrix} g_+ \\ g_- \end{pmatrix} = 0 \qquad (2.11)$$

where $g_\pm = \begin{pmatrix} g_\pm^+ \\ g_\pm^- \end{pmatrix}$. If we write these spinor components as $g_s^\pm(\vec{r}) = R_s^\pm(r)\Theta_s^\pm(\theta)\Phi_s^\pm(\phi)$, where $s$ is the + or − sign, then Eq (2.11) gets separated in all three coordinates as follows

$$\left(\pm\sigma_2 \frac{d}{d\phi} \pm \mathrm{i}\sigma_1 W_\phi\right)\Phi_\pm = \pm\mathrm{i}\sigma_2 \varepsilon_\phi \Phi_\pm, \qquad (2.12\mathrm{a})$$

$$\left(\pm\sigma_1 \frac{d}{d\theta} \pm \mathrm{i}\sigma_2 W_\theta \pm \mathrm{i}\sigma_2 \frac{\varepsilon_\phi}{\sin\theta}\right)\Theta_\pm = \sigma_3 \varepsilon_\theta \Theta_\pm, \qquad (2.12\mathrm{b})$$

$$\begin{pmatrix} 1+\lambda^2 V-\varepsilon & \lambda\sigma_3\left(-\frac{d}{dr}+\frac{\varepsilon_\theta}{r}-\mathrm{i}W_r\right) \\ \lambda\sigma_3\left(\frac{d}{dr}+\frac{\varepsilon_\theta}{r}+\mathrm{i}W_r\right) & -1+\lambda^2 V-\varepsilon \end{pmatrix}\begin{pmatrix} R_+ \\ R_- \end{pmatrix} = 0, \qquad (2.12\mathrm{c})$$

where $\varepsilon_\phi$ and $\varepsilon_\theta$ are the separation constants which are real and dimensionless.

For the case where $W_\phi = 0$ Eq. (2.12a) could be rewritten as $\frac{d}{d\phi}\Phi_\pm = \mathrm{i}\varepsilon_\phi \Phi_\pm$, giving the normalized solution

$$\Phi_s^\pm(\phi) = \frac{1}{\sqrt{2\pi}} e^{\mathrm{i}\varepsilon_\phi \phi}, \qquad (2.13)$$

The requirement that $f_\pm(\phi) = f_\pm(\phi+2\pi)$ puts a restriction on the real values of $\varepsilon_\phi$. Now, $f_\pm(\vec{r}) = \Lambda(\vec{r})g_\pm(\vec{r})$, that is $f_\pm = \frac{\lambda(r,\theta)}{\sqrt{2\pi}} e^{-\frac{\mathrm{i}}{2}\sigma_3 \phi} e^{-\frac{\mathrm{i}}{2}\sigma_2 \theta} \begin{pmatrix} R_\pm^+\Theta_\pm^+ \\ R_\pm^-\Theta_\pm^- \end{pmatrix} e^{\mathrm{i}\varepsilon_\phi \phi}$. Therefore, we obtain the requirement that $e^{\mathrm{i}2\pi\varepsilon_\phi} = -1$. Hence, we should have

$$\varepsilon_\phi = \frac{m}{2}, \qquad m = \pm 1, \pm 3, \pm 5, \ldots \qquad (2.14)$$

Now, Eq. (2.12b) could be rewritten as $\left(\frac{d}{d\theta} - \sigma_3 W_\theta - \sigma_3 \frac{\varepsilon_\phi}{\sin\theta}\right)\Theta_\pm = \mp\mathrm{i}\sigma_2 \varepsilon_\theta \Theta_\pm$. In explicit matrix form it reads

$$\begin{pmatrix} \frac{d}{d\theta} - W_\theta - \frac{\varepsilon_\phi}{\sin\theta} & s\varepsilon_\theta \\ -s\varepsilon_\theta & \frac{d}{d\theta} + W_\theta + \frac{\varepsilon_\phi}{\sin\theta} \end{pmatrix}\begin{pmatrix} \Theta_s^+ \\ \Theta_s^- \end{pmatrix} = 0, \qquad (2.15)$$

where again the sign $s = \pm$. This 2×2 matrix equation decouples, for each angular spinor component, into a second order differential equation as follows

$$\left[\frac{d^2}{d\theta^2} - W_\theta^2 \mp \frac{dW_\theta}{d\theta} - 2\varepsilon_\phi \frac{W_\theta}{\sin\theta} + \varepsilon_\phi \frac{\pm\cos\theta - \varepsilon_\phi}{\sin^2\theta} + \varepsilon_\theta^2\right]\Theta_s^\pm = 0, \qquad (2.16)$$

which resembles the supersymmetric quantum mechanical equation with superpartner potentials $\mathcal{W}^2 \pm \mathcal{W}'$ and eigenvalue $\varepsilon_\theta^2$, where $\mathcal{W} = W_\theta + \frac{\varepsilon_\phi}{\sin\theta}$. The method of supersymmetric quantum mechanics could be used to obtain the solution of this equation.

Nevertheless, we employ here an alternative approach as follows. To simplify the process of obtaining the solution we change variables to the configuration space with coordinate $x \in [-1, +1]$, where $x = \cos\theta$. The integration measure in this space is $\int_0^\pi \sin\theta\, d\theta = \int_{-1}^{+1} dx$. Writing $W_\theta = U(x)/\sin\theta$ casts Eq. (2.16) into the following form

$$\left[ (1-x^2)\frac{d^2}{dx^2} - x\frac{d}{dx} \pm \frac{dU}{dx} - \frac{(U+\varepsilon_\phi)(U+\varepsilon_\phi \mp x)}{1-x^2} + \varepsilon_\theta^2 \right] \Theta_s^\pm = 0. \qquad (2.17)$$

Since the solution of this equation is spanned by $L^2$ functions defined in the compact space with coordinate $x \in [-1, +1]$ then by comparing it with the differential equation of the Jacobi polynomials [4], which are also defined in the same space, we could suggest the following form of solution

$$\Theta_s^\pm(\theta) = A_n (1-x)^\alpha (1+x)^\beta P_n^{(\mu,\nu)}(x). \qquad (2.18)$$

This wavefunction is compatible with the domain of the angular Dirac Hamiltonian and by a proper choice of parameters it could be made to satisfy square integrability and the boundary conditions [5]. $P_n^{(\mu,\nu)}(x)$ is the Jacobi polynomial of order $n$, where $n$ is a non-negative integer. The real dimensionless parameters $\mu, \nu > -1$ and the normalization constant is

$$A_n = \sqrt{\frac{2n+\mu+\nu+1}{2^{\mu+\nu+1}} \frac{\Gamma(n+1)\Gamma(n+\mu+\nu+1)}{\Gamma(n+\mu+1)\Gamma(n+\nu+1)}}. \qquad (2.19)$$

Due to the factor $1/\sqrt{\sin\theta}$ in the spinor $f_\pm(\vec{r})$, which comes from $\lambda(\vec{r})$ and which is equal to $(1-x)^{-\frac{1}{4}}(1+x)^{-\frac{1}{4}}$, then square integrability requires that the real dimensionless parameters $\alpha, \beta \geq \frac{1}{4}$. Substituting (2.18) into Eq. (2.17) and using the differential equation for the Jacobi polynomial we obtain

$$\left\{ \left[ \mu-\nu-2\alpha+2\beta + x(\mu+\nu+1-2\alpha-2\beta) \right]\frac{d}{dx} + \alpha(\alpha-1)\frac{1+x}{1-x} + \beta(\beta-1)\frac{1-x}{1+x} \right.$$
$$\left. -2\alpha\beta + x\left(\frac{\alpha}{1-x} - \frac{\beta}{1+x}\right) \pm \frac{dU}{dx} - \frac{(U+\varepsilon_\phi)(U+\varepsilon_\phi \mp x)}{1-x^2} + \varepsilon_\theta^2 - n(n+\mu+\nu+1) \right\} P_n^{(\mu,\nu)} = 0 \qquad (2.20)$$

The differential equation of the Jacobi polynomial requires that the angular potential function $U(x)$ be linear in $x$. That is, $U(x) = a - bx$, where $a$ and $b$ are real and dimensionless physical parameters. This gives the azimuthal component of the electromagnetic potential as $A_\phi = \frac{a - b\cos\theta}{r\sin\theta}$, which is a combination of Aharonov-Bohm potential whose magnetic flux strength is $2\pi|a-b|$ and a magnetic monopole potential with strength $b$ and singularity along the negative $z$-axis [6]. Requiring that the representation in the solution space, which is spanned by (2.18), be orthogonal dictates that the $x$-dependent factors multiplying $P_n^{(\mu,\nu)}$ and $\frac{d}{dx}P_n^{(\mu,\nu)}$ in Eq. (2.20) must vanish. After some simple, but somewhat lengthy, manipulations we obtain the following results:

$$2\alpha = \mu + \tfrac{1}{2},\ 2\beta = \nu + \tfrac{1}{2}, \qquad (2.21a)$$

$$\mu^2 = \left(a - b + \tfrac{m\mp 1}{2}\right)^2,\ \nu^2 = \left(a + b + \tfrac{m\pm 1}{2}\right)^2, \qquad (2.21b)$$

$$\varepsilon_\theta^2 = \left(n + \tfrac{\mu+\nu+1}{2}\right)^2 - b^2. \qquad (2.21c)$$

The top (bottom) sign in these formulas goes with the corresponding one in the superscript of the angular spinor component $\Theta_s^\pm$. Employing the condition that $\alpha, \beta \geq \frac{1}{4}$ in Eq. (2.21a) gives a stronger constraint on the real values of the parameters $\mu$ and $\nu$ which is that $\mu, \nu \geq 0$. Thus, Eq. (2.21b) gives

$$\mu = \left|a - b + \tfrac{m \mp 1}{2}\right|, \quad \nu = \left|a + b + \tfrac{m \pm 1}{2}\right|. \tag{2.21b'}$$

Separability of the Dirac equation requires that $\varepsilon_\theta$ be the same for the two components of the angular spinor, $\Theta_s^\pm$. Thus, for a given integer $n$ and $m$, the value of $\mu + \nu$ should be the same for both $\Theta_s^+$ and $\Theta_s^-$. That is the value of $\mu + \nu$ is independent of the choice of either the top or bottom signs in (2.21b)'. Now, one can easily show that

$$\mu + \nu = \begin{cases} |2a + m|, & |2a + m| > |2b \pm 1| \\ |2b \pm 1|, & |2a + m| < |2b \pm 1| \end{cases} \tag{2.22}$$

Consequently, simple analysis shows that the separability requirement is satisfied only if either one of the following two conditions is met

$$b = 0, \text{ or, } |2a + m| \geq 2|b| + 1. \tag{2.23}$$

Therefore, in the presence of a magnetic monopole ($b \neq 0$) this condition excludes from the permissible range of values of $m$ the following set of odd integers:

$$m \notin \{-2(|b| + a) - 1 < m < 2(|b| - a) + 1\}. \tag{2.24}$$

Additionally, it is elementary to show that with the expressions (2.21b)' for $\mu$ and $\nu$ one obtains $\mu + \nu \geq 2|b| - 1$ for all real values of $a$ and $b$ and for all integers $m$. Hence, the right hand side of Eq. (2.21c) is always positive and we can choose to write $\varepsilon_\theta^2 = \rho(\rho + 1) + \frac{1}{4}$, where $\rho$ is a real physical parameter. Alternatively, $\varepsilon_\theta = \pm(\rho + \frac{1}{2})$. Therefore, for a given $\rho$ the odd integer $m$ could, in principle, assume any value in the range $m = \pm 1, \pm 3, \pm 5, \ldots, \pm \hat{m}$, where $\hat{m}$ is obtained from Eq. (2.21c) with $n = 0$ as the maximum positive odd integer satisfying

$$\max(\mu + \nu) \leq -1 + 2\sqrt{\left(\rho + \tfrac{1}{2}\right)^2 + b^2}. \tag{2.25a}$$

Using Eq. (2.22) and Eq. (2.23) this condition could be written equivalently as

$$\hat{m} \leq -1 - 2|a| + 2\sqrt{\left(\rho + \tfrac{1}{2}\right)^2 + b^2}. \tag{2.25b}$$

Merging this result with that in (2.23), we conclude that the admissible range of values of the odd integer $m$ when $b = 0$ is $-\hat{m} \leq m \leq \hat{m}$. However, for $b \neq 0$ we should exclude from this range the set of odd integers in (2.24). Now, for any odd integer $m$ in the permissible range obtained above, the non-negative integer $n$ is determined from Eq. (2.21c) as

$$n = \sqrt{\left(\rho + \tfrac{1}{2}\right)^2 + b^2} - \left|a + \tfrac{m}{2}\right| - \tfrac{1}{2} = 0, 1, 2, \ldots \tag{2.26}$$

This is analogous to the spherically symmetric Dirac problem ($a = b = 0$) [1] where $\frac{m}{2} = \pm \frac{1}{2}, \pm \frac{3}{2}, \pm \frac{5}{2}, \ldots, \pm j$ and $n = j - \frac{|m|}{2} = 0, 1, 2, \ldots$ with $j$ being the total angular momentum quantum number ($j = |\ell \pm \frac{1}{2}| = \frac{1}{2}, \frac{3}{2}, \frac{5}{2}, \ldots$). Now, putting all of the above together, we obtain the following angular components of the spinor wavefunction

$$\Theta_s^\pm(\theta) = A_n \sqrt{\sin\theta} \, (1-x)^{\frac{1}{2}\left|a - b + \frac{m \mp 1}{2}\right|} (1+x)^{\frac{1}{2}\left|a + b + \frac{m \pm 1}{2}\right|} P_n^{\left(\left|a - b + \frac{m \mp 1}{2}\right|, \left|a + b + \frac{m \pm 1}{2}\right|\right)}(x), \tag{2.27}$$

where $m = \pm 1, \pm 3, \pm 5, \ldots, \pm \hat{m}$ excluding, for $b \neq 0$, the set (2.24). For each $m$ the non-negative integer $n$ is given by Eq. (2.26). Using the orthogonality relation of the Jacobi polynomials [4], one can easily verify that $\Theta_s^\pm$ as given above makes the angular component of $f_\pm(\vec{r})$ normalizable and orthogonal.

In the following section we complete construction of the total solution space by obtaining the radial component of the spinor wavefunction that solves Eq. (2.12c) in the case where $V(r)$ is the Coulomb potential, $Z/r$, and $Z$ is the electric charge coupling.

### III. SOLUTION OF THE RADIAL DIRAC EQUATION AND THE ENERGY SPECTRUM

Imposing the Lorentz condition by taking $W_r = 0$ and transforming to the four component radial spinor $\begin{pmatrix} R_+ \\ \sigma_3 R_- \end{pmatrix}$ results in a mapping of the radial Dirac equation (2.12c) with the Coulomb potential, $V = Z/r$, into the following

$$\begin{pmatrix} 1 + \lambdabar^2 \frac{Z}{r} - \varepsilon & \lambdabar\left(-\frac{d}{dr} + \frac{\varepsilon_\theta}{r}\right) \\ \lambdabar\left(\frac{d}{dr} + \frac{\varepsilon_\theta}{r}\right) & -1 + \lambdabar^2 \frac{Z}{r} - \varepsilon \end{pmatrix} \begin{pmatrix} R_+ \\ \sigma_3 R_- \end{pmatrix} = 0. \quad (3.1)$$

This matrix wave equation results in two coupled first order differential equations for the two radial spinor components. Eliminating the lower component in favor of the upper gives a second order differential equation. This equation is not Schrödinger-like (i.e., it contains first order derivatives). Obtaining a Schrödinger-like wave equation is desirable because it results in a substantial reduction of the efforts needed for getting the solution. It puts at our disposal a variety of well established techniques to be employed in the analysis and solution of the problem. One such advantage, which will become clear shortly, is the resulting map between the parameters of the relativistic and nonrelativistic problem. This parameter map could be used in obtaining, for example, the relativistic energy spectrum in a simple and straight-forward manner from the known nonrelativistic spectrum. To obtain the Schrödinger-like equation we proceed as follows. A global unitary transformation $e^{\frac{i}{2}\lambdabar\xi\sigma_2}$ is applied to the radial Dirac equation (3.1), where $\xi$ is a real constant parameter. The Schrödinger-like requirement dictates that the parameter $\xi$ should satisfy the constraint $\sin(\lambdabar\xi) = \lambdabar Z/\varepsilon_\theta$, where $-\frac{\pi}{2} \leq \lambdabar\xi \leq +\frac{\pi}{2}$ depending on the signs of $Z$ and $\varepsilon_\theta$. Equation (3.1) will, consequently, be transformed into the following

$$\begin{pmatrix} \frac{\gamma}{\varepsilon_\theta} - \varepsilon + 2\lambdabar^2 \frac{Z}{r} & \lambdabar\left(-\frac{Z}{\varepsilon_\theta} + \frac{\gamma}{r} - \frac{d}{dr}\right) \\ \lambdabar\left(-\frac{Z}{\varepsilon_\theta} + \frac{\gamma}{r} + \frac{d}{dr}\right) & -\frac{\gamma}{\varepsilon_\theta} - \varepsilon \end{pmatrix} \begin{pmatrix} \tilde{R}_+(r) \\ \tilde{R}_-(r) \end{pmatrix} = 0, \quad (3.2)$$

where

$$\gamma = \tau\sqrt{\varepsilon_\theta^2 - \lambdabar^2 Z^2} = \tau\left[\left(n + \tfrac{1}{2}\left|a - b + \tfrac{m\mp 1}{2}\right| + \tfrac{1}{2}\left|a + b + \tfrac{m\pm 1}{2}\right| + \tfrac{1}{2}\right)^2 - b^2 - \lambdabar^2 Z^2\right]^{1/2}. \quad (3.3)$$

$\tau$ is the sign of $\varepsilon_\theta$ (i.e., $\tau = \varepsilon_\theta/|\varepsilon_\theta|$) and

–8–

$$\begin{pmatrix} \tilde{R}_+ \\ \tilde{R}_- \end{pmatrix} = e^{\frac{i}{2}\lambdabar\xi\sigma_2} \begin{pmatrix} R_+ \\ \sigma_3 R_- \end{pmatrix} = \begin{pmatrix} \cos\frac{\lambdabar\xi}{2} & \sin\frac{\lambdabar\xi}{2} \\ -\sin\frac{\lambdabar\xi}{2} & \cos\frac{\lambdabar\xi}{2} \end{pmatrix} \begin{pmatrix} R_+ \\ \sigma_3 R_- \end{pmatrix}. \tag{3.4}$$

Equation (3.3) and $\varepsilon_\theta = \pm(\rho + \frac{1}{2})$ show that real solutions are obtained only if the physical parameters satisfy the constraint that $|\rho + \frac{1}{2}| \geq \lambdabar |Z|$, which could always be satisfied at some nonrelativistic limit defined by a given small enough value of the relativistic parameter $\lambdabar$. Equation (3.2) gives the lower radial spinor component in terms of the upper as follows

$$\tilde{R}_- = \frac{\lambdabar}{\gamma/\varepsilon_\theta + \varepsilon}\left(-\frac{Z}{\varepsilon_\theta} + \frac{\gamma}{r} + \frac{d}{dr}\right)\tilde{R}_+, \tag{3.5}$$

where $\varepsilon \neq -\gamma/\varepsilon_\theta$. Whereas, the resulting Schrödinger-like second order differential wave equation for the upper radial component becomes

$$\left[-\frac{d^2}{dr^2} + \frac{\gamma(\gamma+1)}{r^2} + 2\frac{Z\varepsilon}{r} - \frac{\varepsilon^2 - 1}{\lambdabar^2}\right]\tilde{R}_+(r) = 0. \tag{3.6}$$

For the singular case where $\varepsilon = -\gamma/\varepsilon_\theta$ (i.e., when the energy is negative and at the lower bound of the spectrum), the "kinetic balance relation" (3.5) does not hold. However, the solution is obtained by choosing the inverse of the above global unitary transformation. That is, by taking the negative of the transformation parameter $\xi$ such that $\sin(\lambdabar\xi) = -\lambdabar Z/\varepsilon_\theta$. Then one can show that the required negative energy solution is obtained from the positive energy solution by the map $\varepsilon \to -\varepsilon$, $Z \to -Z$, $\varepsilon_\theta \to -\varepsilon_\theta$, $\tilde{R}_+ \leftrightarrow \tilde{R}_-$. Now, comparing Eq. (3.6) with that of the well-known nonrelativistic Coulomb problem

$$\left[-\frac{d^2}{dr^2} + \frac{\ell(\ell+1)}{r^2} + 2\frac{Z}{r} - 2E\right]\Psi(r) = 0, \tag{3.7}$$

gives, by correspondence, the following map between the parameters of the two problems:

$$Z \to Z\varepsilon, E \to (\varepsilon^2 - 1)/2\lambdabar^2, \ell \to \begin{cases} \gamma & ,\tau > 0 \\ -\gamma - 1 & ,\tau < 0 \end{cases} \tag{3.8}$$

It should be noted that this is a "correspondence" map between the parameters of the two problems and not an equality of the parameters. That is we obtain, for example, the correspondence map $\ell \to \gamma$ but not the equality $\ell = \gamma$. In fact, $\gamma$ is not an integer while, of course, $\ell$ is. Using the parameter map (3.8) in the well-known nonrelativistic energy spectrum of the Coulomb problem, $E_{k\ell} = -Z^2/2(\ell+k+1)^2$ [7], gives the following relativistic spectrum for bound states

$$\varepsilon^+_{k,n,m} = \pm\left[1 + \left(\frac{\lambdabar Z}{k+\gamma+1}\right)^2\right]^{-1/2}, \tau > 0 \tag{3.9a}$$

$$\varepsilon^-_{k,n,m} = \pm\left[1 + \left(\frac{\lambdabar Z}{k-\gamma}\right)^2\right]^{-1/2}, \tau < 0 \tag{3.9b}$$

where $k = 0, 1, 2, ..$ and the dependence on the integers $n$ and $m$ comes form $\gamma$ as given by Eq. (3.3). One can easily see that $\varepsilon^-_{k+1,n,m}\big|_\gamma = \varepsilon^+_{k,n,m}\big|_{-\gamma}$. Therefore, the energy spectrum is two-fold degenerate and could be written collectively as follows:

$$\varepsilon_{k,n,m} = \pm\left[1 + \left(\frac{\lambdabar Z}{k+|\gamma|+1}\right)^2\right]^{-1/2} = \pm\left[1 + \left(\frac{\lambdabar Z}{k+1+\sqrt{\varepsilon_\theta^2 - \lambdabar^2 Z^2}}\right)^2\right]^{-1/2}, \tag{3.10}$$

The only non-degenerate positive energy state is the one associated with the highest energy (upper bound of the spectrum) given by Eq. (3.9b) for $k = 0$, which is equal to $|\gamma/\varepsilon_\theta|$. On the other hand, there exists another non-degenerate negative energy state associated with the lowest energy (lower bound of the spectrum) where $\varepsilon = -|\gamma/\varepsilon_\theta|$. It is obtained from the positive energy solutions by the map shown above Eq. (3.7). Now, the upper radial component $\tilde{R}_+$ of the *positive* energy solution is obtained using the same parameter map (3.8) in the nonrelativistic wavefunction [7]

$$\Psi_{k\ell}(r) = \sqrt{\frac{\omega_{k\ell}\Gamma(k+1)}{\Gamma(k+2\ell+2)}} (\omega_{k\ell} r)^{\ell+1} e^{-\omega_{k\ell} r/2} L_k^{2\ell+1}(\omega_{k\ell} r), \tag{3.11}$$

where $\omega_{k\ell} = -2Z/(k+\ell+1)$ and $Z < 0$. The result is the following radial spinor component which is associated with the positive energy in (3.10) for $k = 0, 1, 2, ..$

$$\tilde{R}_+ = \begin{cases} \sqrt{\frac{\omega_{knm}\Gamma(k+1)}{\Gamma(k+2\gamma+2)}} (\omega_{knm} r)^{\gamma+1} e^{-\omega_{knm} r/2} L_k^{2\gamma+1}(\omega_{knm} r) & ,\tau > 0 \\ \sqrt{\frac{\omega_{knm}\Gamma(k+2)}{\Gamma(k-2\gamma+1)}} (\omega_{knm} r)^{-\gamma} e^{-\omega_{knm} r/2} L_{k+1}^{-2\gamma-1}(\omega_{knm} r) & ,\tau < 0 \end{cases} \tag{3.12}$$

where $\omega_{knm} = -2Z\varepsilon_{knm}/(k+|\gamma|+1)$. The non-degenerate state associated with the highest energy, $\varepsilon = |\gamma/\varepsilon_\theta|$, has the following upper radial spinor component

$$\tilde{R}_+ = \begin{cases} 0 & ,\tau > 0 \\ \sqrt{\frac{2|Z/\varepsilon_\theta|}{\Gamma(-2\gamma)}} \left(2|Z/\varepsilon_\theta|r\right)^{-\gamma} e^{-|Z/\varepsilon_\theta|r} & ,\tau < 0 \end{cases} \tag{3.13}$$

The lower radial component $\tilde{R}_-$ of the positive energy spinor wavefunction is obtained from the upper in (3.12) and (3.13) using the kinetic balance relation (3.5). On the other hand, the radial components of the spinor wavefunction associated with the negative energy solutions are obtained from the positive energy solutions above using the parameter map $\varepsilon \to -\varepsilon$, $Z \to -Z$, $\varepsilon_\theta \to -\varepsilon_\theta$, $\tilde{R}_+ \leftrightarrow \tilde{R}_-$.

The following illustration shows how the elements of the set $\{\Phi(\phi), \Theta^\pm(\theta), \tilde{R}_\pm(r)\}$ cooperate to give a complete and unique specification for each of the four components $\{f_+^\pm, f_-^\pm\}$ of the spinor wavefunction $\psi(\vec{r})$

$$\left.\begin{array}{l} f_+^+ \to \Theta^+ \\ f_+^- \to \Theta^- \end{array}\right\} \tilde{R}_+ \\ \left.\begin{array}{l} f_-^+ \to \Theta^+ \\ f_-^- \to \Theta^- \end{array}\right\} \tilde{R}_- \right\} \Phi \tag{3.14}$$

Explicitly, we obtain the following total wavefunction

$$\psi(\vec{r}) = \begin{pmatrix} if_+ \\ f_- \end{pmatrix} = \begin{pmatrix} i\Lambda g^+ \\ \Lambda g^- \end{pmatrix} = \begin{pmatrix} i\Lambda\begin{pmatrix}\Theta^+\\ \Theta^-\end{pmatrix} R_+ \\ \Lambda\begin{pmatrix}\Theta^+\\ \Theta^-\end{pmatrix} R_- \end{pmatrix} \Phi = \begin{pmatrix} i\Omega\begin{pmatrix}\Theta^+\\ \Theta^-\end{pmatrix}(\eta_+ \tilde{R}_+ - q\eta_- \tilde{R}_-) \\ \Omega\begin{pmatrix}\Theta^+\\ -\Theta^-\end{pmatrix}(q\eta_- \tilde{R}_+ + \eta_+ \tilde{R}_-) \end{pmatrix} \frac{e^{\frac{i}{2}m\phi}}{r\sqrt{2\pi\sin\theta}}, \tag{3.15}$$

where $\Omega = e^{-\frac{i}{2}\sigma_3\phi} e^{-\frac{i}{2}\sigma_2\theta}$, $\eta_\pm = \sqrt{\frac{1}{2}(1 \pm \gamma/\varepsilon_\theta)}$, $q = \text{sign}(Z\varepsilon_\theta) = Z\varepsilon_\theta/|Z\varepsilon_\theta|$ and we have used $\cos(\lambda\xi) = \gamma/\varepsilon_\theta$. In the following section we show that the well-known spherically symmetric result (the Dirac-Coulomb problem) is a special case of our above findings.

Moreover, we obtain the nonrelativistic limit and verify that it agrees with nonrelativistic results reported elsewhere in the literature.

## IV. TWO SPECIAL CASES: SPHERICAL SYMMETRY AND THE NONRELATIVISTIC LIMIT

It is straight-forward to verify that in the spherically symmetric case (where, $a = b = 0$) $\mu + \nu = |m| = 1, 3, 5, ...$ and by using Eq. (2.21c) we obtain

$$\varepsilon_\theta = \pm\left(n + \frac{|m|+1}{2}\right) = \pm 1, \pm 2, \pm 3, ... \tag{4.1}$$

Thus, $\rho = n + \frac{|m|}{2} = \frac{1}{2}, \frac{3}{2}, \frac{5}{2}, ...$ and $\varepsilon_\theta$ becomes the spin-orbit quantum number which is usually referred to by the symbol $\kappa$

$$\varepsilon_\theta = \kappa = \pm\left(j + \tfrac{1}{2}\right) = \begin{cases} +j+\tfrac{1}{2} &, \ell = j+\tfrac{1}{2} \\ -j-\tfrac{1}{2} &, \ell = j-\tfrac{1}{2} \end{cases}, \tag{4.2}$$

where $\ell$ is the orbital angular momentum quantum number and $j$ is the total angular momentum (orbital plus spin), which is equal to $\rho$. Moreover, we can write $n = \rho - \frac{|m|}{2} = j - \frac{|m|}{2} = 0, 1, 2, ...$ Thus,

$$\tfrac{m}{2} = -j, -j+1, ..., j-1, j. \tag{4.3}$$

This range of values of $m$ is also obtainable using Eq. (2.25b). By substituting these results in (3.10) we recover the familiar relativistic energy spectrum for the Dirac-Coulomb problem [1,8]

$$\varepsilon_{k,j} = \pm\left[1 + \lambdabar^2 Z^2 \Big/ \left(k + 1 + \sqrt{(j+\tfrac{1}{2})^2 - \lambdabar^2 Z^2}\right)^2\right]^{-1/2}. \tag{4.4}$$

The total angular component of the spinor wavefunction could now be written as

$$Y_s^\pm(\theta, \phi) = \frac{\sqrt{\Gamma(j+\tfrac{1}{2}|m|+1)\Gamma(j-\tfrac{1}{2}|m|+1)}}{\sqrt{2^{|m|+1}\pi}\,\Gamma(j+\tfrac{1}{2})} \sqrt{\sin\theta}\,(1-x)^{\frac{|m\mp 1|}{4}}(1+x)^{\frac{|m\pm 1|}{4}} P_{j-\tfrac{1}{2}|m|}^{\left(\frac{|m\mp 1|}{2}, \frac{|m\pm 1|}{2}\right)}(x)\,e^{\tfrac{i}{2}m\phi}, \tag{4.5}$$

where we have used the identity

$$\Gamma\left(n + \tfrac{|m\mp 1|}{2} + 1\right)\Gamma\left(n + \tfrac{|m\pm 1|}{2} + 1\right) = \Gamma\left(n + \tfrac{|m|}{2} + \tfrac{1}{2}\right)\Gamma\left(n + \tfrac{|m|}{2} + \tfrac{3}{2}\right)$$
$$= \left(n + \tfrac{|m|}{2} + \tfrac{1}{2}\right)\Gamma\left(n + \tfrac{|m|}{2} + \tfrac{1}{2}\right)^2 = \left(j+\tfrac{1}{2}\right)\Gamma\left(j+\tfrac{1}{2}\right)^2 \tag{4.6}$$

The radial component is obtained from (3.12) and (3.13) with $\gamma = \tau\sqrt{\kappa^2 - \lambdabar^2 Z^2}$ and $\varepsilon_\theta = \kappa$.

Finally, by taking the nonrelativistic limit ($\lambdabar \to 0$), the positive energy spectrum in (3.10) becomes

$$\varepsilon_{knm} \cong 1 - \lambdabar^2 \frac{Z^2}{2(k+|\varepsilon_\theta|+1)^2}. \tag{4.7}$$

Since in this limit $\varepsilon \to 1 + \lambdabar^2 E$, where $E$ is the nonrelativistic energy, then the non-relativistic spectrum is obtained as

$$E_{knm} = -Z^2 \Big/ 2\left(k + |\varepsilon_\theta| + 1\right)^2$$
$$= -Z^2 \Big/ 2\left[k + 1 + \sqrt{\left(n + \tfrac{1}{2}\left|a - b + \tfrac{m \mp 1}{2}\right| + \tfrac{1}{2}\left|a + b + \tfrac{m \pm 1}{2}\right| + \tfrac{1}{2}\right)^2 - b^2}\right]^2 \quad (4.8)$$

This agrees with the nonrelativistic spectrum obtained in the literature for a charged particle in the Coulomb plus the Aharonov-Bohm potential of magnetic flux strength $2\pi a$ (with $b = 0$) [9] or in the presence of a magnetic monopole of strength $b$ and A-B magnetic flux $2\pi|a - b|$ [10]. When comparing our results with those in the literature one should note that the azimuth phase quantum number in most of those publications is to be identified not with the odd integer $m$ above but with $\frac{m \pm 1}{2} = 0, \pm 1, \pm 2, \dots$.

**ACKNOWLEDGMENTS**

I am grateful to O. Mustafa for the hospitality at the Physics and Chemistry Departments of the Eastern Mediterranean University, North Cyprus, where part of this work was carried out. The help provided by M. S. Abdelmonem in literature search is highly appreciated.